\documentclass[floatfix,prl,aps,showpacs,superscriptaddress]{revtex4}
\usepackage{epsfig,graphicx,tabularx}
\usepackage{subfigure}
\usepackage{amsmath,amsfonts,amssymb}
\usepackage{color,braket}
\usepackage{siunitx}
\usepackage{soul}
\usepackage{ulem}

\begin{document}

\title{Hidden Magnetism in Periodically Modulated One Dimensional Dipolar Fermions} 
\author{S. Fazzini} \affiliation{Institute for condensed matter physics and complex systems, DISAT, Politecnico di Torino, I-10129, Italy}
\author{A. Montorsi} \affiliation{Institute for condensed matter physics and complex systems, DISAT, Politecnico di Torino, I-10129, Italy}

\author{M. Roncaglia}\affiliation{Institute for condensed matter physics and complex systems, DISAT, Politecnico di Torino, I-10129, Italy}

\author{L. Barbiero} 
\affiliation{CNR-IOM DEMOCRITOS Simulation Center and SISSA, Via Bonomea 265, I-34136 Trieste, Italy}
\affiliation{Dipartimento di Fisica e Astronomia "Galileo Galilei", Universit\`a di Padova, 35131 Padova, Italy}

\begin{abstract}
The experimental realization of time-dependent ultracold lattice systems has paved the way towards the implementation of new Hubbard-like Hamiltonians. We show that in a one-dimensional two-components lattice dipolar Fermi gas the competition between long range repulsion and correlated hopping induced by periodically modulated on-site interaction allows for the formation of hidden magnetic phases, with degenerate protected edge modes.  The magnetism, characterized solely by string-like nonlocal order parameters, manifests in the charge and/or in the spin degrees of freedom. Such behavior is enlighten by employing Luttinger liquid theory and numerical methods.  The range of parameters for which hidden magnetism is present can be reached by means of the currently available experimental setups and probes.
 \end{abstract}


\maketitle



Since the Haldane's discovery of a gapped phase with no local order in 1983 \cite{haldane}, hidden magnetic orders \cite{NiRo} have attracted huge interest. In this context two very recent experiments involving organic molecular compounds \cite{janani}  and an oxide of nickel spin chain \cite{xu} have obtained relevant results. Nowadays cold atomic systems offer an ideal platform to simulate fundamental quantum physics \cite{Bloch2008}. Indeed proposals for the realization of hidden charge magnetism \cite{dallatorre,dalmonte,kobayashi, cohen, pohl} have been carried out. Meanwhile the possible realization of interaction induced hidden spin orders in fermionic systems is still an unexplored scenario.\\
At the same time, investigations of periodically modulated quantum systems \cite{eckardt1} have predicted very interesting effects \cite{eckardt, greschner, greschner2}. They have stimulated impressive experimental achievements like frustrated classical magnetism \cite{struck}, gauge potentials \cite{struck2}, ferromagnetic domains \cite{parker} and the realization of new particle-hole symmetric Hubbard-like Hamiltonians with correlated hopping processes (CHPs) \cite{meinert}. The latter are believed to be responsible for fundamental still open questions \cite{montorsi_libro}, one of these being the celebrated $\eta$-superconductivity \cite{yang}.\\
A configuration closer to real materials \cite{Lahaye2009} can be realized in trapped ultracold atomic systems with strong long-range dipolar interaction, like magnetic atoms \cite{Griesmaier2005,Lu2011,Aikawa2012} and polar molecules \cite{Ni2008,Wu2012, Takekoshi2014}. In case of $Er$ magnetic atoms, this research line has produced the recent experimental realization \cite{baier} of a paradigmatic model in condensed matter, the extended Bose-Hubbard model. Furthermore out-of-equilibrium dipolar systems have been both used to generate quantum magnetic Hamiltonians \cite{Yan2013,DePaz2013} and proposed to study disorderless many-body localized regimes \cite{me}.

Motivated by the aforementioned reasons, in this paper we investigate the properties of a dipolar fermionic mixture subject to a rapid time periodic modulation of the on-site interaction and trapped in a one-dimensional (1D) optical lattice. In this regime Floquet theory can be applied. It allows to derive an effective time independent model where an additional term of CHPs appears. When we treat the effective model within bosonization approach \cite{giamarchi}, its behavior is reduced to that of two spin-charge separated sine-Gordon models. The latter turns out to capture well the charge sector, predicting in particular the presence of hidden charge order; and to only partly describe the behavior in the spin sector, since Haldane spin order appears to be ruled out. In fact, once quasi-exact density matrix renormalization group (DMRG) \cite{white} calculations are performed, a further spin gapped region is found with respect to bosonization predictions. Noticeably, this is characterized by the presence of hidden magnetic order in the spins. This magnetism can be solely detected by the non-vanishing of string-like nonlocal order parameters (NLOPs).  Finally we discuss how all our achievements can be experimentally reproduced with the ongoing experimental setups involving magnetic atoms.    

\begin{figure}[t]
	\includegraphics[scale=0.4]{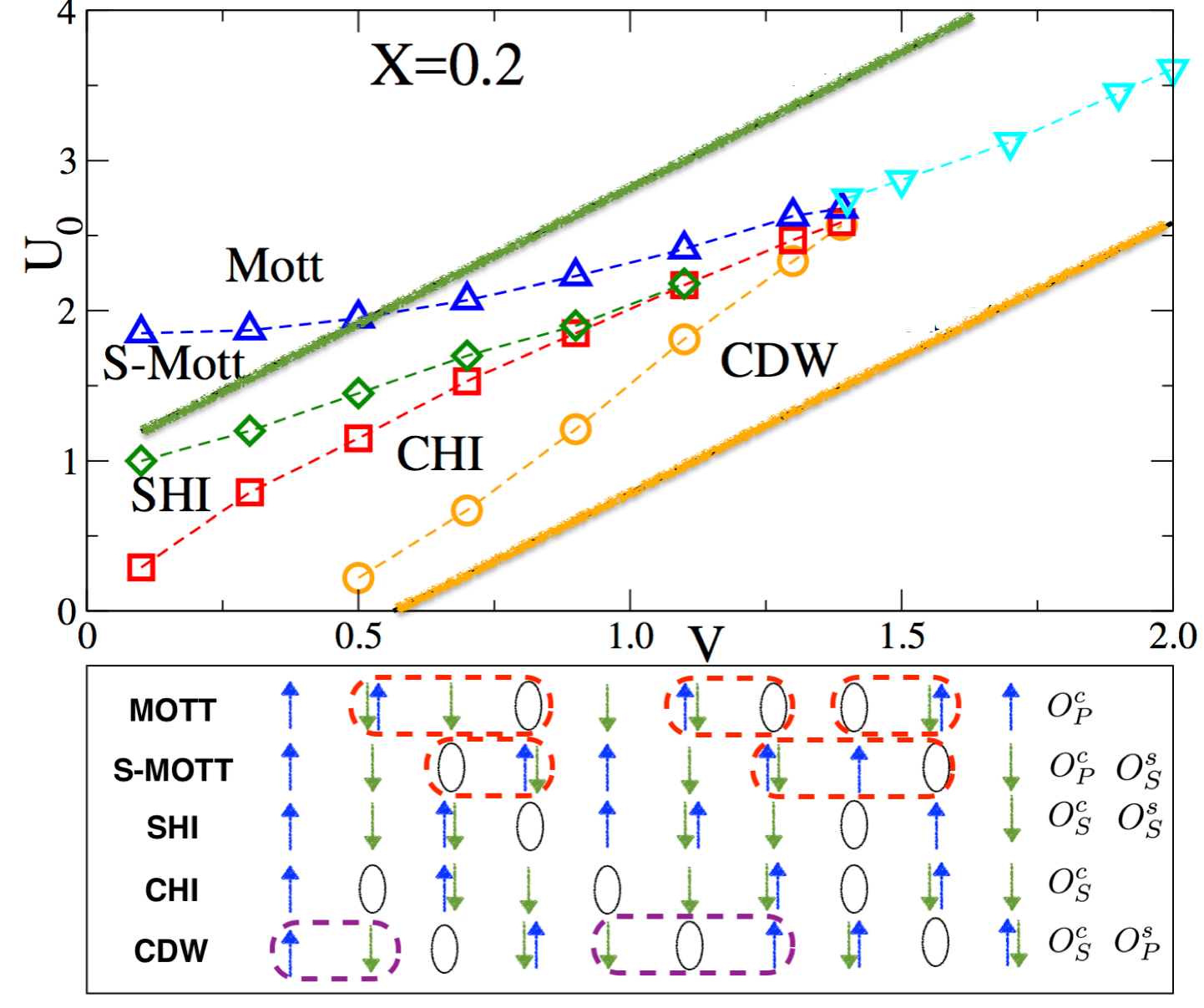}
	\caption{(Color online) \textit{Upper panel}: DMRG (symbols) and bosonization (solid lines) phase diagram of (\ref{EFHeff}) as a function of $U_0$ and $V$ with $J=1$ and $X=0.2$. \textit{Lower panel}: Cartoon of the phases with the relative NLOPs. The red (purple) dashed circles show the doublon-holon (spin up-down) virtual excitations}
	\label{fig1}
\end{figure}

\paragraph{Model.}
We consider a balanced unit density two components ($\sigma=\uparrow,\downarrow$) dipolar Fermi mixture of $N$ particles \cite{note1} with onsite periodically modulated interaction trapped in a 1D optical lattice. Within a single band approximation, i. e. for a deep optical lattice, the extended
Fermi-Hubbard model \cite{Lahaye2009} gives an accurate description of the system
\begin{eqnarray}
H=&-&J\sum_{\langle ij \rangle}\sum_{\sigma=\uparrow,\downarrow}(c^\dagger_{i\sigma}c_{j\sigma}+h.c.)\nonumber\\
&+&U(t)\sum_j n_{j\uparrow}n_{j\downarrow}+V\sum_{j,r\geq 1}\frac{n_jn_{j+r}}{r^3}, 
\label{EFH}
\end{eqnarray}
where $\langle..\rangle$ denotes nearest neighbors,
$c_{j\sigma}$ ($c_{j\sigma}^{\dagger}$) destroys~(creates) a $\sigma$-fermion in the $j$-th site of a lattice of length $L$ and $n_j=\sum_\sigma n_{j,\sigma}$ counts the total number of particles at site $j$. Crucially in cold atomic experiments all the couplings, namely the hopping rate $J$, the onsite interaction $U$ and the long range dipolar repulsion $V$ may be independently controlled by modifying the lattice depth, the transversal confinement
\cite{Goral2003, bartolo}, using Feshbach resonances, and/or controlling the
orientation and strength of the polarizing field. The time dependence in (\ref{EFH}) can be induced by a rapid variation of the scattering length \cite{rapp} producing a periodic modulation of the form $U(t)=U_0+U_1\cos(\omega t)$ which consequently makes $H(t)=H(t+T)$ being $T=2\pi/\omega$. In the regime $\omega>>U_0/\hbar,J/\hbar$, Floquet theory can be used \cite{grifoni} to approximately remove the time dependence. Indeed, analogously to the $V=0$ case \cite{diliberto}, we find that this kind of interaction modulation generates an effective time independent Hamiltonian where the hopping processes are renormalized by the density, namely the CHPs
\begin{eqnarray}
H_{eff}=&-&J\sum_{\langle ij \rangle,\sigma}\left(c^\dagger_{i\sigma}c_{j\sigma}+h.c\right)\Big(1-X(n_{i\bar{\sigma}}-n_{j\bar{\sigma}})^2\Big)\nonumber\\
&+&U_0\sum_j n_{j\uparrow}n_{j\downarrow}+V\sum_{j,r>0}\frac{ n_j n_{j+r}}{r^3}
\label{EFHeff}
\end{eqnarray}
where $X=1-{\cal J}_0(U_1/\hbar\omega)$ is the CHPs rate, ${\cal J}_0$ is the first kind Bessel function, and  $\bar{\sigma}$ denotes the other component with respect to $\sigma$. The model (\ref{EFHeff}) in the $V=0$ regime has attracted huge interest in the context of cuprate superconductors \cite{arrachea} while the $X=0$ case has been studied in the context of time independent dipolar fermions, see \cite{didio} and references therein. Meanwhile only few partial analysis have tried to approach the full $H_{eff}$ \cite{me1, nakamura, japaridze}, in case of just nearest-neighbor repulsion. From the other side, our Fig. \ref{fig1} shows how the model (\ref{EFHeff}) presents a rich phase diagram including quantum regimes with hidden magnetic properties and topological order. As explained in the following sections, the latter phases can be characterized solely by means of non-local order parameters.  

\paragraph{Nonlocal Order Parameters.}
In the context of lattice fermions a very fundamental role is played by NLOPs of parity- and string-like form. Their correlation functions can be written respectively as
\begin{equation}
O_P^{\nu}(r)=\langle e^{\imath\pi\sum_{j<r}S_j^{\nu}}\rangle 
\label{parity}
\end{equation}
\begin{equation}
O_S^{\nu}(r)=\langle S_l^{\nu}e^{\imath\pi\sum_{l\leq j<l+r}S_j^{\nu}}S_{l+r}^{\nu}\rangle \quad ,
\label{string}
\end{equation}
where $\nu=c,s$ refers to the charge and spin degrees of freedom, and the charge and spin operators are defined as: $S_j^c=(1-n_j)$ and $S_j^s=(n_{j\uparrow}-n_{j\downarrow})$. The relevance of NLOPs lies in the fact that they act as order parameters for gapped 1D phases \cite{NiRo,dallatorre,montorsi} without breaking any continuous symmetry, thus in agreement with the Mermin-Wagner theorem \cite{mermin}. Also, non-vanishing NLOPs characterize \cite{monprep} symmetry protected topological phases \cite{wen2,pol} obtained by group cohomology. In particular a finite $O_S^\nu$ in the thermodynamic limit is a signature of a phase with non-trivial topological properties, noticeably the presence of degenerate edge modes \cite{monprep,keselman}, with charge or spin fractionalization. These facts have motivated their intensive use to study 1D fermionic systems \cite{me1,montorsi,dutta,dhar,deb} helping to display physical properties not captured by the usual two-point correlation functions. More precisely, in fermionic systems the role of $O_P^\nu$ is to signal the presence of trivial Mott- (BEC-) like orders with virtual excitations consisting of correlated pairs of holon-doublon (for $\nu=c$) or single electrons with up-down spin (for $\nu=s$) \cite{montorsi}. Meanwhile $O_S^\nu$ captures hidden non-trivial ``dilute" Haldane-like antiferromagnetic orders \cite{me1,haldane} of holon-doublon ($\nu=c$) or up-down spins ($\nu=s$).

\paragraph{Dynamics vs Effective Model.}

\begin{figure}[h]
\includegraphics[scale=0.5]{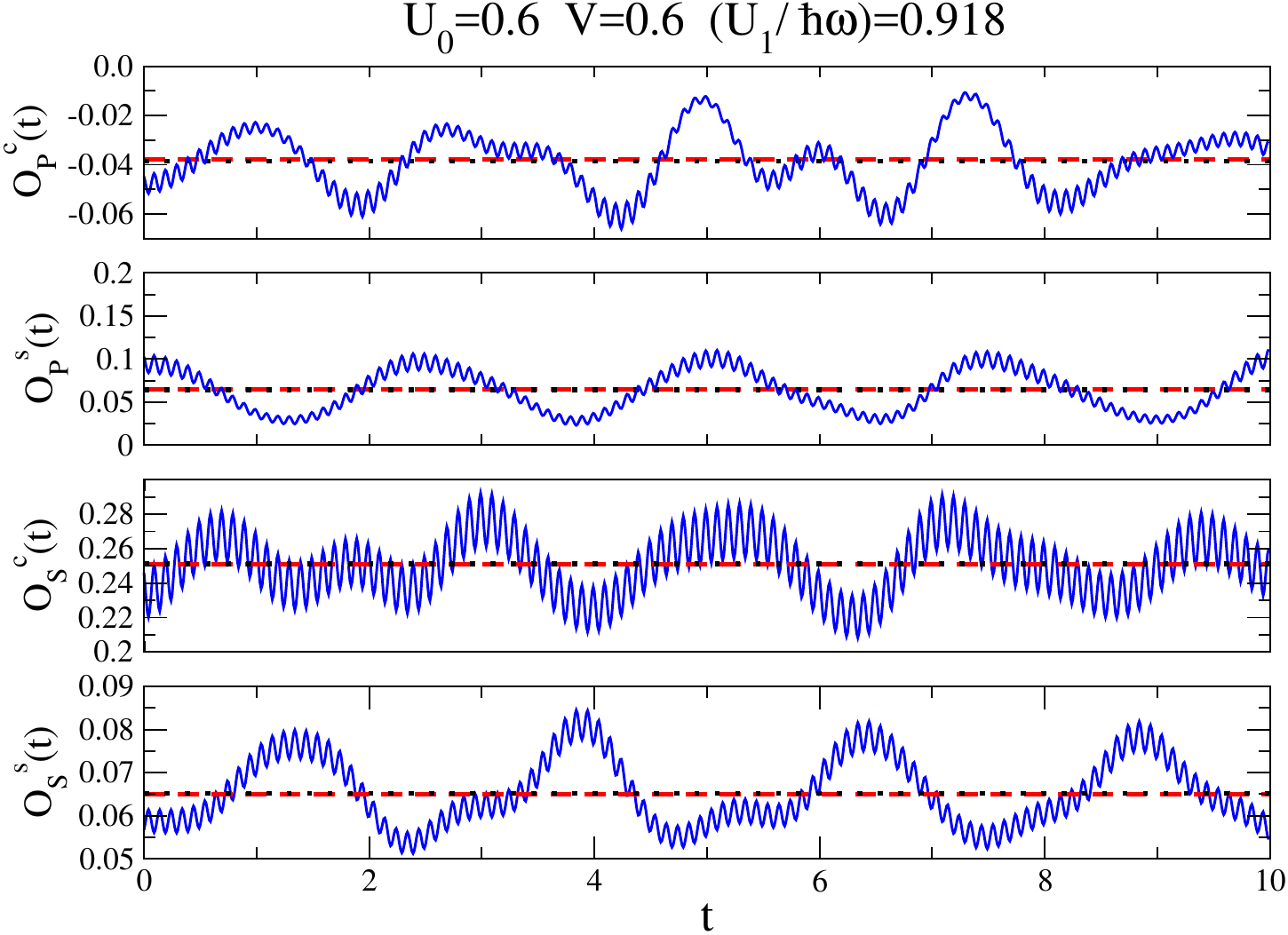}
\caption{(Color online) Blue continuous lines are the time evolution of NLOPs, red dashed lines are the time-averaged values of NLOPs and black dotted lines are the NLOP values given by $H_{eff}$ eq. (\ref{EFHeff}). The extrapolated value of the strings is $O_S^{\nu}(L/2)$ while the extrapolated value of the parities is $(O_P^{\nu}(L/2)+O_P^{\nu}(L/2+1))/2$. All the results refer to a system of $L=8$ sites and both couplings and inverse time are expressed in unit of $J$. The amplitude of the correlated hopping processes in the static model is $X=0.2$.}
\label{fig2}
\end{figure}

In order to check the validity of the Floquet theory we compare the finite size NLOP values obtained by simulating both the time dependent model (\ref{EFH}) and the derived time independent effective model (\ref{EFHeff}). The time-dependent simulations are performed by means of Exact Diagonalization starting from the initial ground state with couplings $J,U_0,V$ and at $t>0$ a time periodicity $U(t)$ is applied in order to get CHPs of strength $X=0.2$. After that, we monitor the time evolution and we evaluate the time-averages of the NLOPs \cite{note}. As clearly shown in Fig. \ref{fig2} all the time averaged NLOP values are in good agreement with the ones obtained by exactly diagonalizing (\ref{EFHeff}) with couplings $J,U_0,V,X$ \cite{note2}. The latter result allows us to safely study $H_{eff}$ in order to get the full phase diagram of (\ref{EFH}) without losing some fundamental aspect encoded in the time dependence.   

\begin{figure}[t]
\includegraphics[scale=0.5]{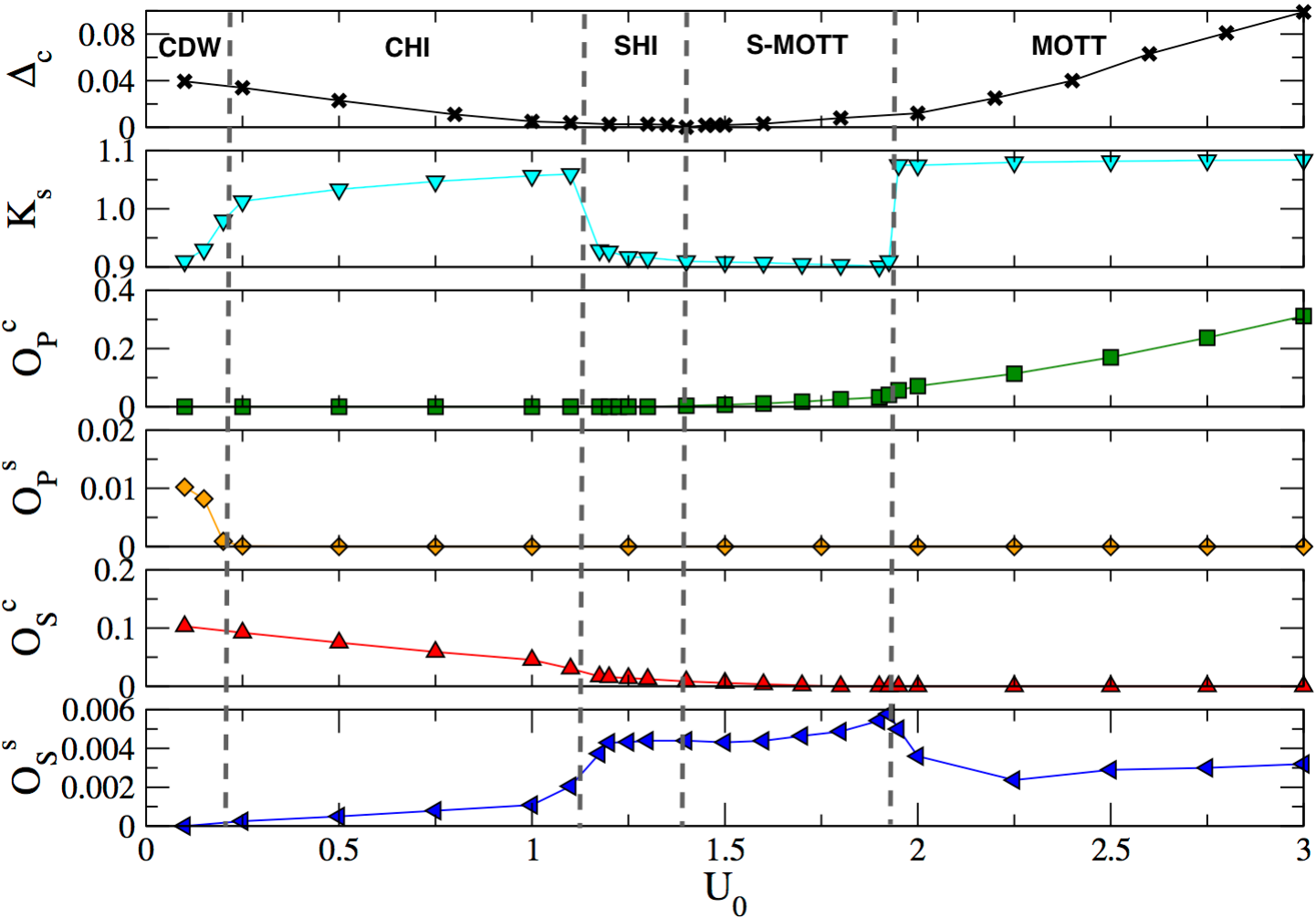}
\caption{(Color online) Thermodynamic limit of charge gap, spin Luttinger constant and NLOPs of (\ref{EFHeff}) for $J=1$, $V=0.5$ and $X=0.2$ as a function of $U_0$. $\Delta_c$ is extrapolated by using open boundary conditions and sizes up to $L=44$. $K_s$ and the NLOPs are extrapolated by using periodic boundary conditions and sizes up to $L=36$. The extrapolated value of the strings is $O_S^{\nu}(L/2)$ while the extrapolated value of the parities is $(O_P^{\nu}(L/2)+O_P^{\nu}(L/2+1))/2$. In all our DMRG simulations we cut $r$ to three nearest neighbors keeping up to 1600 DMRG states and performing up to 6 finite size sweeps.}
\label{fig3}
\end{figure}

\begin{figure}[t]
	\includegraphics[scale=0.6]{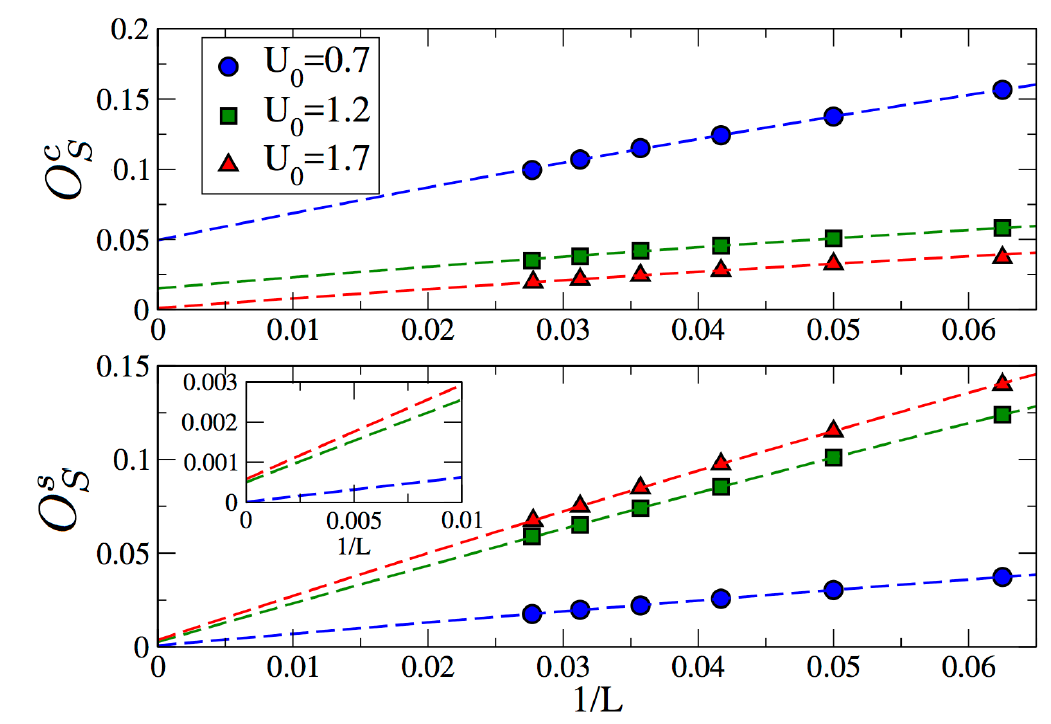}
	\caption{(Color online) \textit{a)} Finite size extrapolation of $O_s^c$. \textit{b)}Finite size extrapolation of $O_s^s$. \textit{Inset} Zoom of the fit in \textit{b)}. All the data are obtained by fixing $J=1$, $X=0.2$ and $V=0.5$ for different values of $U_0$ and lattice sizes $L$ ranging form 16 to 36. The extrapolated value of the strings is $O_S^{\nu}(L/2)$. In all our DMRG simulations we cut $r$ to three nearest neighbors keeping up to 1600 DMRG states and performing up to 6 finite size sweeps.}
	\label{figstring}
\end{figure}

\paragraph{Luttinger Liquid Analysis.}
Here we report the results of a bosonization analysis \cite{giamarchi} of $H_{eff}$, which details can be found in \cite{suppl}. The Hamiltonian can be regarded as the sum of three contributions: ${\cal H}={\cal H}_{c}+{\cal H}_{s}+{\cal H}_{cs}$. In the weak coupling limit, each of the first two contributions has the form of a sine-Gordon model in the $\nu$ sector, namely 
\begin{equation}
{\cal H}_{\nu}=\int dx\left[H_{0\nu}+\frac{m_{\nu}v_{\nu}}{2\pi a^{2}}\cos(\sqrt{8\pi}\phi_{\nu})\right]
\label{H_sG}
\end{equation}	
with $H_{0\nu}=\frac{v_{\nu}}{2}\Big[K_{\nu}(\partial_{x}\theta_{\nu})^{2}+\frac{1}{K_{\nu}}(\partial_{x}\phi_{\nu})^{2}\Big]$. The contribution ${\cal H}_{cs}=\frac{M_{cs}}{\pi a}\int dx \cos(\sqrt{8\pi}\phi_c)\cos(\sqrt{8\pi}\phi_s)$, which couples the spin and the charge sectors, is irrelevant in a renormalization group analysis, having scaling dimension 4 and therefore it is usually neglected.
The massive phases of the two decoupled sine-Gordon models can be analyzed in the asymptotic limit by studying the renormalization group flow equations. In $\nu$ sector the transition line to a gapped phase is determined by
the competition between the kinetic and the mass terms, which generates the equation
	\begin{equation}
	2(K_{\nu}-1)< |m_{\nu}| \quad .
	\label{RG}
	\end{equation}
The sign of $m_{\nu}$ causes the field $\phi_{\nu}$  in eq. (\ref{H_sG}) to pin to one of the two values $0$ or $\pm \sqrt{\pi/8}$, which correspond to the appearance of two distinguished nonlocal orders, namely the parity and the Haldane string orders. In fact, in the continuum limit the NLOPs (\ref{parity}) and (\ref{string}) become \cite{me1}: $O_P^\nu\rightarrow \langle \cos\sqrt{2\pi}\phi_\nu\rangle^2$, and $O_S^\nu\rightarrow\langle \sin\sqrt{2\pi}\phi_\nu\rangle^2$. This observation also allows to connect a non-vanishing Haldane string order with the presence of degenerate edge modes typical of a non-trivial symmetry protected topological phase. Indeed, within the bosonization framework, they are observed at the edge between the trivial ($\phi_{\nu}=0$) and the non-trivial ($ \phi_{\nu}=\pm\sqrt{\pi/8}$) phases  \cite{monprep,keselman}, and are characterized by fractional charge (for $\nu=c$) or spin (for $\nu=s$).
The solution of inequality (\ref{RG}) in the charge sector produces the following transition line
\begin{equation}
U_{0c}= \frac{3}{2} \zeta(3) V+\frac{16}{\pi} X \quad . 
\label{U_0c}
\end{equation}	
For $V\neq 0$, it marks the boundary between two gapped phases, which can be distinguished by means of different NLOPs. Indeed, as $U_0$ crosses this critical point from higher to lower values, the topological nature of the insulating state changes from trivial ($O_P^c\neq 0$) to non-trivial ($O_S^c\neq 0$).\\
On the other hand, it is found that the spin sector is gapless if $U_0\geq U_{0s}$, where
	\begin{equation}
	U_{0s}= \frac{3}{2} \zeta(3) V-\frac{16}{\pi} X \quad .
	\label{U_0s}
	\end{equation}
Instead, the gapped phase obtained for $U_0< U_{0s}$ is characterized by a parity order: $O_P^s\neq 0$.\\
Summarizing, we find that bosonization analysis predicts the presence of three insulating phases, separated by the two transition lines (\ref{U_0c}) and (\ref{U_0s}), as illustrated in Fig. \ref{fig1}, where the bosonization results are represented by the green and yellow solid lines. For $U>U_{0c}$, the state is characterized by finite $O^c_P$, thus configuring as a Mott insulating phase with trivial topological properties. 
Instead, between the two solid lines, i.e., for $U_{0s}<U_0<U_{0c}$, the system is ordered by $O_S^c$, which identifies a charge Haldane insulator (CHI). Finally, for $U<U_{0s}$ the charge Haldane order coexists with a spin parity order, thus designating the presence of a charge density wave (CDW) locally ordered phase \cite{me1}.\\
In the end, we would like to stress that,
since the $su(2)$ spin invariance of the Hamiltonian imposes constraints on the coefficients of the sine-Gordon model in the spin channel, within the one loop bosonization analysis no phases with hidden spin order can be present. In other words, the approach described in this section is not able to predict the presence of possibly existing phases with $O_S^s\neq 0$. In fact, as we will see in the next section, the numerical analysis shows the evidence of this order for limited regions inside both the Mott and the CHI phases. The bosonization approach can be improved by releasing the requirement of spin-charge separation and by including the effect of higher order harmonics.
\paragraph{DMRG results.}
The bosonization analysis is expected to give reliable results in the weak coupling regime. Below we perform a further analysis based on quasi exact DMRG simulations to explore the full phase diagram.
Since an insulating behavior is expected everywhere except along a critical line, as a first step we evaluate the thermodynamic limit of the charge gap $\Delta_c=(E(N+2)+E(N-2)-2E(N))/2$ being $E(N)$ the ground state energy of $N$ particles. The result is displayed in the upper panel of Fig. \ref{fig3} for fixed $X=0.2$ and $V=0.5$. It  clearly show that $\Delta_c$ vanishes only in one point thus signaling a continuous phase transition between two insulating phases. DMRG results become more crucial when analyzing the behavior in the spin degrees of freedom, where CHPs are known to make bosonization less predictive \cite{bosspin}. Here, as a first step we evaluate the thermodynamic limit of the Luttinger spin constant $K_{s}=\lim_{q\rightarrow 0} \pi S^s(q)/q$ where $S^s(q)=1/L\sum_{k,l}e^{\imath q(k-l)}\left(\langle S^s_kS^s_l\rangle-\langle S^s_k\rangle\langle S^s_l\rangle\right)$ is the spin structure factor. Luttinger liquid theory predicts $K_{s}=1(0)$ in absence  (presence) of a finite spin gap, defined as the energy variation in flipping one spin. Here both logarithmic corrections and finite size effects make very difficult to get sharp $0,1$ values. Nevertheless a well established and accurate approximation, see \cite{sengupta}, is to consider a spin gapless (gapped) phase in presence of $K_{s}>1(<1)$: the transition point is then fixed by the crossing of the value 1. As clearly visible in Fig. \ref{fig3} the analysis based on the thermodynamic limit of $K_s$ surprisingly finds, for small values of $V$, a further large spin gapped phase ranging in a region around the single point where $\Delta_c=0$.  The above results allow us to identify all the gapped regions of the phase diagram, as reported in Fig. \ref{fig1}. The nature of each phase can be better characterized by studying the behavior of the different nonlocal order parameters (see bottom panels of Fig. \ref{fig3}). Based on this analysis, for large $V$, we find a phase with CDW ($O_S^c,O_P^s\neq 0$) order, as expected, in analogy with the nearest neighbor extended Fermi-Hubbard model (see for instance \cite{nakamura}). The similarities extend also to the large $U_0$ region where a charge gapped Mott phase signaled by $O_P^c$ is present. Between the CDW and Mott regions instead, the fully gapped phase with bond ordering (BOW) --characteristic of the extended Fermi-Hubbard model and signaled by a non-zero value of both parity operators-- is destroyed by the CHPs.
Indeed Fig. \ref{fig3} shows that, at intermediate $U_0,V$, three different phases characterized by hidden magnetism take place. In particular as predicted by bosonization, a phase having as an order parameter only $O_S^c$ appears (CHI). The latter reproduces in a two-species fermionic system the same charge hidden antiferromagnetic order of the well known topological Haldane phase studied in the context of spin-1 chains \cite{haldane}, extended Bose-Hubbard model \cite{rossini,deng,batrouni}, and multicomponent fermions \cite{nonne}. Moreover, at variance with the bosonization results, Fig. \ref{fig3}  also shows that both the Mott and CHI regions are partially replaced by fully gapped phases, due to the presence of the spin string order ($O_S^s\neq 0$) coexisting with the charge order, meaning that $\uparrow$ and $\downarrow$ spins are alternated and diluted in an arbitrary number of holons and doublons properly organized. In particular, at fixed $V$ and by increasing $U_0$, we first find a phase with the two strings being simultaneously non-vanishing ($O_S^c,O_S^s\neq 0$), thus describing hidden magnetism in both degrees of freedom: holons and doublons are themselves diluted and alternated. This is called spin Haldane insulator (SHI). By a further increase of the onsite interaction, the hidden antiferromagnetic charge order ($O_S^c$) is replaced by the Mott-like charge order ($O_P^c$), where holons and doublons are organized in localized pairs. Meanwhile $O_S^s$ remains finite in a further range, thus giving rise to spin gapped Mott phase (S-Mott). In Fig. \ref{figstring}, we show the finite size extrapolation of the string order parameters for different values of the onsite interaction strength, namely in the CHI, SHI and S-Mott phases.\\
The full phase diagram is shown if Fig. \ref{fig1}, where the numerical results (symbols) are compared with the bosonization predictions (solid lines). Moreover, in the lower panel a schematic picture of the phases is drawn to help the reader understand their connotations in terms of nonlocal orders.
\begin{figure}[h]
\includegraphics[scale=0.5]{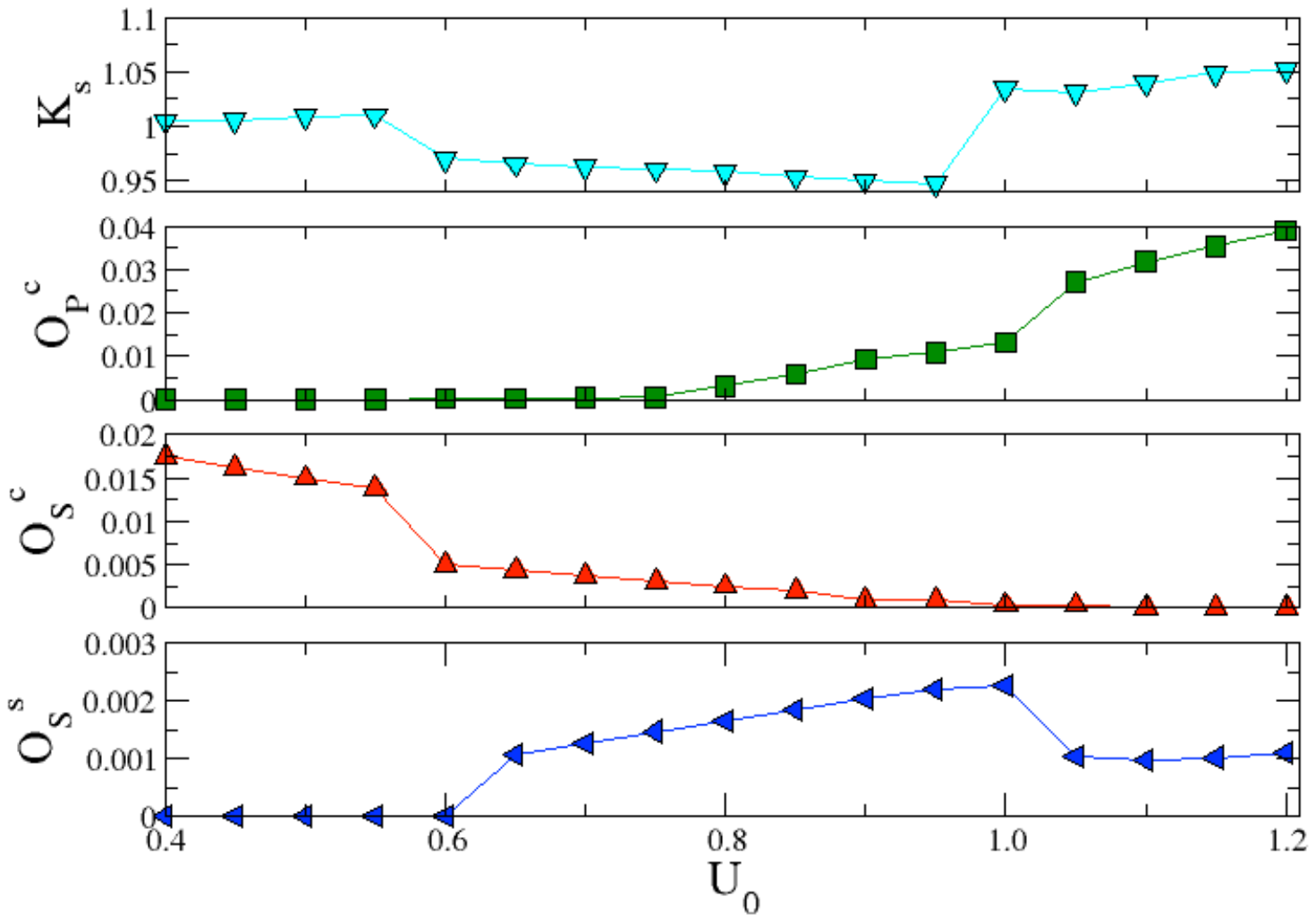}
\vspace*{-0.5cm}
\caption{(Color online) Thermodynamic limit of spin Luttinger constant $K_s$ and NLOPs for $J=1$, $V=0.3$ and $X=0.05$ as a function of $U_0$. $K_s$ and the NLOPs are extrapolated by using periodic boundary conditions and sizes up to $L=24$. The extrapolated value of the strings is $O_S^{\nu}(L/2)$ while the extrapolated value of the parities is $(O_P^{\nu}(L/2)+O_P^{\nu}(L/2+1))/2$. In all our DMRG simulations we cut $r$ to three nearest neighbors keeping up to 1200 DMRG states and performing up to 6 finite size sweeps.}
\label{fig1suppmat}
\end{figure}
We have checked that all our phases with hidden magnetism are stable and robust with respect to varying $X$. In particular, increasing $X$, the shape of the region with finite $O_S^s$ is preserved,  whereas the size of the region with charge string order increases, thus giving rise to an even bigger SHI regime. The crucial point, as shown in Fig. \ref{fig1suppmat} is that the $O_S^s$ order persists also for very weak X, meaning that CHPs are the solely responsible for these kinds of hidden spin orders.  
\paragraph{Discussion} The above DMRG results envisage a new scenario where hidden spin string order is explicitly induced by interaction without breaking the full $su(2)$ spin symmetry of the Hamiltonian, at variance with previous results \cite{keselman, kainaris}.  Moreover, both the SHI and the S-Mott phases configure as fully gapped phases with no local order, at variance with bosonization predictions.\\
In fact one may further notice that the presence even for weak coupling of spin gapped phases not predicted by one-loop bosonization was already observed for a similar model \cite{bosspin}. In that case the inclusion of higher order harmonics in the spin channel of bosonization analysis was subsequently proved \cite{AD} to induce the opening of the observed spin gapped phase for $\phi_s=0$. However it must be stressed that here the further spin gapped phase is non-trivial, i.e. it opens in correspondence of $\phi_s=\pm \sqrt{\pi/8}$. This suggests that the spin-charge coupling term ${\cal H}_{cs}$ should have not been neglected. Otherwise the $su(2)$ symmetry of the sine-Gordon model would imply $K_s<1$ only for $m_s<0$, which solution describes just the trivial phase. In fact, following an approach similar to \cite{nakamura}, the role of ${\cal H}_{cs}$ in a charge gapped phase can be understood by considering in the spin channel an effective sine-Gordon model with renormalized mass $m_s^*=m_s+M_{cs}<\cos\sqrt{8\pi}\phi_c>$, where $\phi_c$ fixed to $0$ (S-Mott) or $\pm\sqrt{\pi/8}$ (SHI). Depending on the sign of $M_{cs}$, one sees that in this case even for a negative $m_s$, $m_s^*$ can become positive: in particular, this happens in presence of Haldane charge order iff $M_{cs}<0$ (SHI phase), whereas for trivial charge order one must have $M_{cs}>0$ (S-Mott phase). The same approach could also be applied to justify the shift which appears in the CHI-CDW transition line with respect to bosonization predictions, exploring the case $m_s^*<0$.

\paragraph{Experimental realization.} 
The previous quantum phases could be studied by using a mixture of Erbium isotopes. In particular fermionic ${}^{167}Er$ \cite{aikawa} as well as bosonic ${}^{168}Er$ \cite{baier} isotopes are currently available in laboratories. The scattering length of the ${}^{168}Er$ can be accurately tuned to reach a practically hard-core regime, thus giving rise to an effective two components Fermi mixture \cite{note3}. At $\ang{30}\leq\theta\leq\ang{90}$ ($\theta$ being the angle between the orientation of the dipoles and the interparticles distance,) a recoil energy $E_R=h\times4.3KHz$, and an appropriate lattice depth should allow to easily achieve the values $0.5\lesssim V/J\lesssim 2$ which is exactly the regime where hidden magnetism is predicted. Feshbach resonance to tune ${}^{167}Er$-${}^{168}Er$ onsite interaction should become available \cite{frisch} and, in order to get CHPs, a rapid time dependent modulation can be applied following the procedure in \cite{meinert}. Performing measurements in a dynamically environment is very challenging. Nevertheless in a recent experiment \cite{bordia} involving periodically modulated fermions, local correlations have been probed. This can be done by a sudden frozen of the system and subsequently using techniques used in static configurations where NLOPs have been already experimentally measured \cite{endress,hilker}.



\paragraph{Summary.} 
We have shown that periodical onsite modulation of lattice dipolar fermions allows to realize Hamiltonians with long range dipolar interaction and correlated hopping processes terms. These drive the system from the static configuration of the extended Hubbard model to states with hidden magnetism. The latter appears in the charge, spin or both sectors, and can be detected solely by NLOPs. Our findings pave the way towards the study of interaction induced hidden magnetic orders and their non-trivial topological effects, such as the formation of degenerate edge modes. The phases can be detected with the currently available experimental setups and probes.



{\it Acknowledgements.--}  We thank L. Dell'Anna, F. Meinert and J. Simonet for discussions. This work was supported by MIUR~(FIRB 2012, Grant No. RBFR12NLNA-002). L. B. thanks the CNR-INO BEC Center in Trento for CPU time and the DISAT at Politecnico di Torino for the hospitality.



\end{document}